\documentclass[prl,aps,twocolumn,showpacs,superscriptaddress,groupedaddress]{revtex4-1}\usepackage{graphicx}
\usepackage{microtype}
\usepackage{amsmath}
\usepackage{amssymb}
\usepackage{textgreek}
\usepackage[some]{background}

\SetBgScale{1}
\SetBgContents{\parbox{20cm}{%
  \Huge Preprint of article published in Biological Cybernetics\\[18.5cm]\rotatebox{180}{\Huge }}}
\SetBgColor{black}
\SetBgAngle{270}
\SetBgOpacity{0.2}

\begin{document}
\title{A Fundamental Inequality Governing the Rate Coding Response of Sensory Neurons}
\author{Willy Wong}
\email{willy.wong@utoronto.ca}
\affiliation{Department of Electrical and Computer Engineering, University of Toronto, Toronto M5S3G4}
\affiliation{Institute of Biomedical Engineering, University of Toronto, Toronto M5S3G9}
\date{\today}

\begin{abstract}
A fundamental inequality governing the spike activity of peripheral neurons is derived and tested against auditory data.  This inequality states that the steady-state firing rate must lie between the arithmetic and geometric means of the spontaneous and peak activities during adaptation.  Implications towards the development of auditory mechanistic models are explored.
\end{abstract}
\maketitle
 \BgThispage
\section{Introduction}
Sensory coding concerns how the nervous system encodes information about physical stimuli \cite{kandel2013principles}.  This paper explores the encoding of intensity information by the rate activity of sensory neurons.  Within this topic, a number of important issues in theoretical neurophysiology are also addressed.  

Models in neurophysiology are generally required to show some degree of compatibility with experimental data before publication in scholarly journals.  However their success must be tempered by the following considerations.  The first is the number of parameters.  While there are different ways to model neurons \cite{herz2006modeling}, many models invariably require a large number of parameters.  This is particularly evident for realistic or conductance-based models e.g. \cite{almog2016realistic,wang2022multimodal}, although so-called simplified models can also have large numbers of parameters depending on the type of phenomena it describes \cite{brunel2010modeling}.  If the parameter values are not known \textit{a priori} this can lead to a problem of overfitting.  von Neumann has been attributed to saying: ``With four parameters I can fit an elephant, and with five I can make him wiggle his trunk.'' \cite{dyson2004meeting}  A high dimensional model can exhibit a wide range of quantitative behaviour.  Not only can the parameters not be determined uniquely \cite{achard2006complex,gutenkunst2007universally,o2015computational}, but compatibility with data is not necessarily proof of correctness of the model.  To deal with overfitting, one can make use of large datasets.  However this is not always possible as experimental data can be limited.  

A second concern is that theory in neurophysiology often lacks new predictions of a quantitative nature that can be falsified experimentally \cite{gamez2012baconian}.  Most neurophysiological models, particularly of the phenomenological kind, are created to account for existing observations.  Ensuring compatibility with known experimental results is easier than predicting new, yet-to-be observed phenomena.  This is perhaps the most difficult hurdle for any potential theory of neurophysiology.

In response to these challenges, this paper seeks to address the following issues.  The first is the derivation of an inequality governing the rate activity of sensory neurons which is believed to be obeyed universally across different modalities and animal species.  The inequality constitutes a new prediction by a theory of sensory coding that was most recently advanced in \cite{wong2020rate}.  The inequality can be falsified from experimental data and requires no knowledge of any of the parameters. Thus the veracity of both the prediction and the theory can be demonstrated without the selection or fine tuning of parameters.  The paper also clarifies the relationship between the inequality and a previously derived result known as the ``geometric mean law'' \cite{wong2020rate,wong2021consilience}.  Finally, while the inequality is derived without consideration of physiological mechanism, it can aid in the exploration of mechanisms underlying sensory transduction.  An illustration is provided with respect to auditory mechanistic modelling.

We begin by reviewing a theory of sensory information processing from which the inequality is derived.

\section{An information theoretical approach to peripheral processing}
There have been ongoing efforts to understand the rate coding of sensory neurons in terms of measurement uncertainty or entropy  \cite{norwich1977information,norwich1993information,norwich1995universal,wong1997physics,wong2013perceptual,wong2020rate}.  The idea behind the theory is straightforward: The sensory receptor samples the stimulus repeatedly and averages these samples.  By the central limit theorem, the uncertainty in the sample mean is normally distributed when sample size is large.  Calculating the Shannon entropy yields an equation relating entropy to the variance of the measurements as well as the sample size.  Three final steps are required to obtain an equation governing the rate response: The association of variance with the mean of the stimulus via a power function; sample size approaches its optimal value by a first-order relaxation process; and firing rate is set equal to the measurement uncertainty or entropy.  This last equation in particular seeks to change the very nature of how we understand the sensory encoding of intensity \cite{norwich1993information}. 

Written out in full, the equations are
\begin{align}
&F=kH \label{gut1}\\
&H=\dfrac{1}{2}\log\left(1+\dfrac{\beta \left(I+\delta I\right)^p}{m}\right) \label{gut2}\\
&\dfrac{dm}{dt}=-a (m-m_{eq}) \label{gut3}\\
&m_{eq}=(I+\delta I)^{p/2} \label{gut4}
\end{align}
where $F$ is the firing rate of the peripheral neuron, $H$ the entropy, $I$ the stimulus magnitude and $t$ the duration of stimulation.  $m$ is the sample size and $m_{eq}$ its optimal or equilibrium value.  The remaining variables are parameters and are described briefly.  The proportionality constant $k$ has units of spikes per second; $\beta$ has units rendering the argument of the logarithm in (\ref{gut2}) dimensionless.  The exponent $p$ is derived from the Tweedie family of probability distributions and relates variance to a power function of the mean.  The value of $p$ is restricted to be greater than equal to one, and should be derivable from the statistics of the sensory stimulus, i.e. in vision $p=1$ due to Poisson statistics.  $\delta I$ is internal noise with the same units as the stimulus magnitude $I$. Finally, $a$ is the inverse time constant of neural response with units of Hz.

Equations (\ref{gut1}-\ref{gut4}) were developed without consideration of underlying mechanism, derived in full and tested successfully against a number of experimental conditions \cite{wong2020rate}.  They can be used to solve the time-varying response of the neuron $F$ given any time-varying input $I$ by following these steps.  First, $I$ is parameterized as a function of time $t$, allowing for the calculation of the optimal sample size $m_{eq}$.  From this, $m$ is solved from the differential equation.  Finally, inserting both $m$ and $I$ into $H$ and $F$ yields the predicted neural response.  The numerical evaluation of the response requires knowledge of the five positive parameters $k, \beta, p, \delta I, a$ which are typically obtained by curve-fitting the equations to experimental data.  However, precise determination of the parameter values can be difficult as the equations are prone to overfitting.

%
Neural adaptation describes the phenomenon that occurs when a unit is initially left unstimulated and in equilibrium. A stimulus is introduced and held constant for $t \ge 0$, and the response of the unit can be described by three phases: the spontaneous activity before the stimulus onset, the peak activity that occurs during or shortly after the stimulus presentation, and the steady-state activity that is observed after the adaptation process is completed \cite{benda2014spike}.  From the theory, we can solve for the adaptation response as follows.  A constant level of stimulus implies that the optimal sample size $m_{eq}$ in (\ref{gut4}) is constant.  From this, we solve for the sample size $m$ as a function of time from (\ref{gut3}) to obtain
\begin{equation}
m(t)=m(0)e^{-at}+m_{eq}\left(1-e^{-at}\right) \label{solution}
\end{equation}
Due to the continuity condition of (\ref{gut3}), $m(0)$ equals the value of $m$ just prior to the stimulus being turned on.  For $t \le 0$, we have $I=0$ and the unit is in steady-state.  Therefore, $m(0)=\delta I^{p/2}$.  Substituting $m$ into (\ref{gut2}) and (\ref{gut1}) gives the following for the adaptation response,
\begin{align}
F&(I,t) = \nonumber \\
&\, \frac{1}{2}k\log\left(1+\dfrac{\beta \left(I+\delta I\right)^p}{\delta I^{p/2}e^{-at}+\left(I+\delta I \right)^{p/2}\left(1-e^{-at}\right)}\right) \label{adaptation}
\end{align}
From here, we identify the expressions for the spontaneous, peak and steady-state activities to be $\text{SR}=F(0,\infty)$, $\text{PR}=F(I,0)$ and $\text{SS}=F(I,\infty)$ and obtain
\begin{align}
& \text{SR} = \tfrac{1}{2}k\log\left(1+\beta \delta I^{p/2}\right) \label{SR} \\
& \text{PR} = \tfrac{1}{2}k\log\left(1+\beta \left(I+\delta I\right)^p/\delta I^{p/2}\right) \label{PR} \\
& \text{SS} = \tfrac{1}{2}k\log\left(1+\beta \left(I+\delta I\right)^{p/2}\right) \label{SS}
\end{align}
These expressions will be used in the next section.

The solution of (\ref{gut1})-(\ref{gut4}) to inputs of various types (step, square wave, sinusoidal, ramp, etc) compare well to experimental measurements and are detailed in \cite{wong2020rate}.  The equations can also be solved numerically using just a few simple lines of computer code (see Appendix A for a numerical solution for the adaptation response).  There are, however, several drawbacks to the theory.  The predicted response falls short in two key aspects: (a) There is no saturation of response at high intensities; (b) The predictions are purely deterministic and do not account for the stochastic nature of the response.

\section{An inequality governing the peripheral activity}
Next, we derive an inequality governing the steady-state activity of a sensory neuron based on the fixed points of the adaptation function.

We begin by defining $x=\beta\delta I^{p/2}$ and $y=\beta\left(I+\delta I \right)^p/\delta I^{p/2}$, and also set $k/2=1$ as it has no bearing on the final result.  Note that $0 \le x \le y$ because all of the parameters are positive and $I \ge 0$.  The spontaneous SR, peak PR and steady-state activities SS from (\ref{SR})-(\ref{SS}) can be rewritten as
\begin{align}
& \text{SR} = \log\left(1+x\right) \\ 
& \text{PR} = \log\left(1+y\right) \\
& \text{SS} = \log\left(1+\sqrt{xy}\right)
\end{align}

From here, we show that the steady-state activity obeys the inequality
\begin{equation}
\sqrt{\text{PR} \times \text{SR}} \le \text{SS} \le \frac{\text{PR}+\text{SR}}{2} \label{inequality}
\end{equation}
That is, the steady-state activity is bounded by the geometric and arithmetic mean of the spontaneous and peak activities.  This result is independent of intensity or the choice of any of the parameters in the equation $(k, \beta, p, \delta I, a)$.  This surprising result constrains the amount of adaptation that occurs in a sensory neuron by limiting the steady-state activity to lie within the mean of the peak and spontaneous activities.

To prove (\ref{inequality}), we first establish the upper bound for SS:
\begin{align*}
\frac{\text{PR}+\text{SR}}{2}& = \frac{\log\left( 1+x+y+xy \right)}{2} \\
&\ge  \frac{\log\left( 1+2\sqrt{xy}+xy \right)}{2} \\
&=\log\left(1+\sqrt{xy} \right) =\text{SS}
\end{align*}
where we have used $\sqrt{xy} \le \left(x+y\right)/2$ for $x,y \ge 0$.  

The lower bound requires the observation that if $\log f(z)$ is a twice differentiable function that is concave with respect to $u=\log z$ then
\begin{align*}
\log f\left(e^{\frac{u_1+u_2}{2}}\right) &\ge \frac{\log f\left(e^{u_1}\right)+\log f\left(e^{u_2}\right)}{2} \\
&= \log \sqrt{f\left(e^{u_1}\right)f\left(e^{u_2}\right)}
\end{align*}
or, in terms of $z$, we have
\begin{equation*}
f\left(\sqrt{z_1z_2} \right) \ge \sqrt{f\left(z_1\right)f\left(z_2\right)} \label{concave}
 \end{equation*}
These equations are obtained by generalizing the usual definition of concavity with both the function and its argument transformed.  Choosing $f(z)=\log\left(1+z \right)$ provides the inequality needed for (\ref{inequality}). That is, 
\begin{align*}
\text{SS}&=\log\left(1+\sqrt{xy} \right) \\
&\ge \sqrt{\log\left(1+x \right)\log\left(1+y \right)}  = \sqrt{\text{PR} \times \text{SR}}
\end{align*}
To check the concavity of $\log f(z)$, we evaluate the second derivative of  $\log \log\left(1+e^{u} \right)$ with respect to $u$:
\begin{equation*}
\frac{d^2\log f\left(e^{u}\right)}{du^2} = \frac{e^u\left[\log\left(1+ e^{u}\right)-e^{u}\right]}{\left(1+e^{u} \right)^2\log^2\left(1+e^u\right)} \le0
\end{equation*}
where the inequality is due to $\log(1+z)\le z$ for $z=e^u \ge0$.  Thus $\log \log(1+z)$ is concave with respect to $\log z$.  

Both the lower and the upper bounds are now proved, and (\ref{inequality}) is established in full.  There is also a weaker inequality that can be obtained from (\ref{inequality}), but one that does not require knowledge of SR.  Since $\text{SR} \ge 0$, we conclude trivially that $\text{SS} \le \text{PR}/2$.  That is, the steady-state activity cannot exceed one half of the peak activity.  Finally, in the absence of stimulus, both PR and SS equal SR and the inequality shows the correct limiting behaviour for $I=0$.

\section{Results and Discussion}
Equation (\ref{inequality}) is a result of considerable mathematical beauty which, if true, highlights the underlying simplicity of the workings of a sensory neuron.  Next we test the inequality against experimental data.  Since the theory from which the inequality is derived is free from the consideration of mechanism, (\ref{inequality}) is expected to work across different modalities and animal species.  However, this paper will examine specifically testing over auditory data due to the precise nature of adaptation experiments in hearing. For example, sound signals can be controlled to a high degree of accuracy, making hearing an ideal modality for repeated measurements. Traditionally, modalities such as temperature, taste, and smell lack the same level of stimulus control.  Adaptation testing in vision can also be challenging due to lateral contributions from other units, hindering the ability to probe single unit responses \cite{enroth1973adaptation}. Additionally, adaptation testing requires a sudden stimulus onset which hearing can tolerate relatively well.

%
There is, however, one additional reason for using auditory data.  While (\ref{inequality}) can be tested against individual measurements of adaptation (i.e. a single adaptation response will yield three values which can be compared to the predictions of the inequality), a single measurement is less satisfying than \textit{multiple measurements from the same unit}.  Since (\ref{inequality}) is independent of intensity, one can obtain a range of results by measuring adaptation under different intensities.  In hearing, there are a number of studies that have carried out such experiments.

\begin{figure*}[!h]
\begin{center}
\includegraphics[width=.9\textwidth]{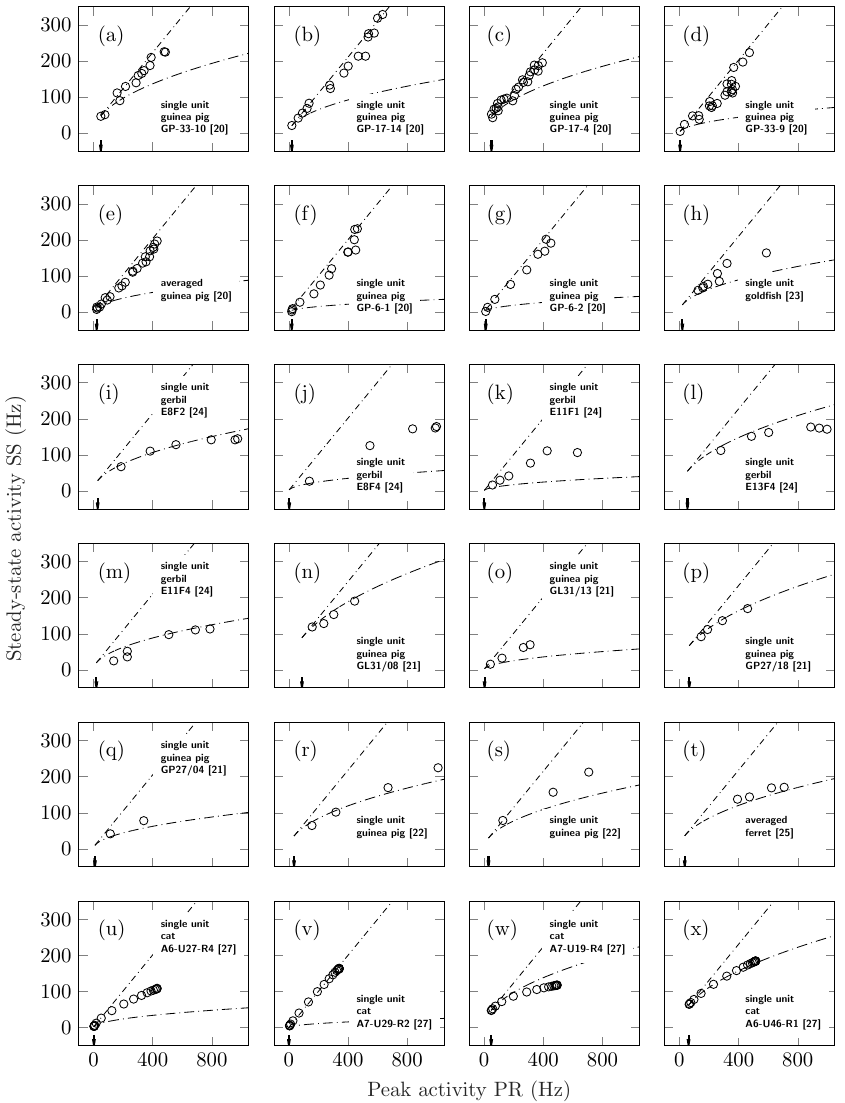}
\caption{Steady-state activity plotted versus peak activity for a number of studies.  In all panels, the dashed lines show the theoretical upper and lower bounds of (\ref{inequality}).  No fitted parameters were required to plot these bounds.  SS vs PR from (a-g) single or averaged guinea pig fibre recordings (figs. 11a-d, 12, 17a and 4a from \cite{smith1975short}); (h) saccular nerve fibres of goldfish (fig. 3 from \cite{fay1978coding}); (i-m) single fibre gerbil recordings (figs. 4 and 5 from \cite{westerman1984rapid}); (n-q) single guinea pig fibre recordings (figs. 1 and 2 from \cite{yates1985very}); (r-s) single guinea pig fibre recordings (figs 3a and 3b from \cite{muller1991relationship}); (t) averaged ferret data (fig. 6 from \cite{sumner2012auditory}); (u-x) single cat fibre recordings (figs. 12e-h from \cite{peterson2021simplified}).  The spontaneous activity of each unit is indicated by an arrow pointing towards the x-axis.}\label{fig1}
\end{center}
\end{figure*}

Fig. \ref{fig1} shows the results of several auditory studies plotting steady-state (SS) versus peak activity (PR) over different intensities together with the theoretical bounds predicted by (\ref{inequality}).  \textit{It is important to remember that there are no fitted parameters required to plot these bounds.}  All that is needed is knowledge of SR which can be obtained directly from the data measurements.  One should be mindful that the bounds are not statistical in nature as they are derived from deterministic equations and are only expected to be satisfied on average.  Fig. \ref{fig1} includes all studies of peripheral measurements of auditory adaptation measured across different intensities known to the author (see Methods section for study inclusion criteria).  With the exception of panels (e) and (t), each panel is generated from a series of measurements from a single unit.  These studies encompass measurements from a number of animal species including guinea pigs (a-g, n-q, r-s) \cite{smith1975short,yates1985very,muller1991relationship}, fish (h) \cite{fay1978coding}, gerbils (i-m) \cite{westerman1984rapid}, ferrets (t) \cite{sumner2012auditory} and cats (u-x) \cite{heil2011improved,peterson2021simplified}.  

With few exceptions, the agreement between data and theory is outstanding, with the data approaching both the upper and lower limits consistently.  In the case where the agreement is less than perfect (see particularly panels i, l and w), the data tends to fall below the lower bound.  Panel (v) among others is worth highlighting.  In this case, the spontaneous activity is low enough to take $\text{SR} \approx 0$.  As such, (\ref{inequality}) simplifies to $\text{SS} \le \text{PR}/2$, and SS is restricted to lie below the diagonal.  Finally, it is unclear how the peak vs steady-state curves will behave when the neuron's response begins to saturate at high intensities; there may very well be violations to the inequality. While systematic deviations do not appear in Fig. \ref{fig1}, very few of the studies probe the saturation response.

There is also evidence that (\ref{inequality}) is obeyed widely across other modalities and animal species.  Taking the lower bound as an \textit{approximation} for the steady-state activity, we have $\text{SS}=\sqrt{\text{PR} \times \text{SR}}$.  That is, the inequality is now replaced by a single equation which determines the value of SS exactly.  This equation was first derived in \cite{wong2020rate} and compared extensively with data drawn from a wide range of studies from different modalities (e.g. proprioception, touch, taste, hearing, vision and smell) and across animal species of different phyla (chordata, arthropoda, mollusca and cnidaria) \cite{wong2021consilience}.  In total, the ``geometric mean law'' was tested over 250 measurements of adaptation and found to hold to good approximation.  A lack of systematic testing in the other modalities preclude an analysis similar to Fig. \ref{fig1}.  A double-log plot was used to check conformance to this law: A plot of SS vs PR should yield a straight line with slope equalling 1/2.  However, the actual value of slope, when calculated across studies pooled together in aggregate, was found to exceed the predicted value.  This is entirely consistent with (\ref{inequality}) in that the geometric mean is in fact the lower bound.  Real data is expected to grow at a rate equal to or exceeding a square root function.  However, it remains unknown whether this relationship holds for all types of units, particularly those found in active sensory systems like electroreception or temperature sensing, or for fast-adapting phasic units \cite{wong2020rate,wong2021consilience}.

One might wonder if there exists a simpler explanation of the inequality without requiring an elaborate theory of sensory coding.  One possibility is that the results are an artifact of the method by which the firing rates are being determined.  Spike frequencies are difficult to estimate when the spiking probability is time-varying.  Certain methods have been proposed to deal with time-varying rates (e.g. \cite{shimazaki2007method}); however, adaptation poses a special challenge due to the rapidly changing firing frequency.  The calculated value of the peak rate PR, in particular, can vary depending on the positioning of the count bins, stimulus rise time e.g. \cite{heil2015basic}, duty cycle of the stimulus \cite{rhode1985characteristics,relkin1991recovery}, as well as the width of the window used to calculate the firing rate.  However, it should be noted that in two thirds of the panels in Fig. \ref{fig1}, the rates were extracted by the original study authors themselves; see panels (a)-(g), (i)-(m) and (u)-(x).  Errors of a systematic variety are also unlikely since the data follows both the upper and lower bounds.  Moreover, it would be quite a coincidence if the error matches the predictions of a theory which has been under development for nearly fifty years, and had nothing to do with the data collected in Fig. \ref{fig1}.  Ultimately (\ref{inequality}) stands as it is -- a theoretical prediction in neurophysiology.  If the existing data is found to be lacking, the prediction can help guide future experiments of higher precision.

If we accept that the data obeys the inequality, then what gives rise to these bounds?  The answer is not hard to find.  The firing rate equation takes on the mathematical form $\log\left(1+x\right)$ where $x$ has dependency on intensity.  This expression has asymptotic behaviour of either $x$ or $\log x$.  From (\ref{SS}), we observe that this gives rise to either a power or a log law, both of which have been used extensively to describe the growth of firing rate with intensity. This in part leads to the upper and lower bounds in Fig. \ref{fig1}.

\section{Implications for mechanisms of sensory transduction}
The inequality was derived from a theory that is free from the consideration of mechanism.  However, non-mechanistic models can work with mechanistic models to help further the understanding of sensory transduction.  This section is devoted to bridging between these two approaches with particular application to audition.

\subsection{The inequality and auditory transduction}
A number of mechanistic models have been proposed to describe the process of generating the peripheral auditory spike response, e.g. \cite{eggermont1973analog,schroeder1974model,smith1982adaptation,meddis1986simulation,westerman1988diffusion,heil2011improved}.  For an extensive review of past models see \cite{hewitt1991evaluation,zhang2005analysis,peterson2021simplified}. There are also more extensive end-to-end computational systems which attempt to model the entire process from sound to spike response.  Such examples include the MATLAB Model of Auditory Periphery (MAP) \cite{meddis2011matlab} and its earlier version, the Development System for Auditory Modelling (DSAM) \cite{omard2021dsam}, as well as the developments that comprise the Zilany-Bruce-Carney model (among other contributors; henceforth referred to as ZBC) \cite{carney1993model,heinz2001auditory,zilany2014updated,bruce2018phenomenological}.  These systems take sound inputs and generate the associated animal-specific spike rate response from auditory neurons.  Both are computational auditory systems: a collection of models from various contributors which can account for the outer, middle and inner ear characteristics as well as the inner hair cell-auditory nerve synapse to model the process of neurotransmitter release determining spike activity.  In the Supplementary Information, simulations using the default settings of each simulator together with a comparison to the inequality (\ref{inequality}) are detailed.  While the simulation results are found to obey the inequality, this should not come as a surprise: these systems were developed using the very same data found in Fig. \ref{fig1}.  

At the heart of each system is a model governing the transduction between physical stimulation and spike generation.  We will briefly review four phenomenological models, two of which form the core of the MAP/DSAM and the ZBC systems.  Coverage of these models is superficial.  There is no detailed analysis beyond a simple cursory explanation, nor are definitions provided for the many parameters or variables.  The interested reader is referred to the original publications.  The goal is simply to survey these approaches, and to get to the underlying equations which can then be solved for the peripheral response.  Some discussion is then devoted towards the use of (\ref{inequality}) to further develop these models.

\subsection{Implications for models without mechanism of intensity coding or mechanism of adaptation}
The first is a model proposed in \cite{westerman1984rapid} (hereby referred to as the Westerman-Smith or WS model) which forms a core part of the ZBC computational system.  The WS approach involves a two-compartmental model of transmitter release in the inner hair cell synapse of the auditory nerve fibre.  The use of the model within ZBC was later modified to include fractional noise and power-law adaptation dynamics \cite{zilany2009phenomenological,zilany2014updated}, and most recently includes a synaptic vesicle docking model \cite{bruce2018phenomenological}.  However, the core idea lies with the original WS model.  The two compartment model includes both a local and immediate compartment: 
\begin{align}
&V_I \dot{c_I}\left(t\right)=-p_I c_I\left(t\right) + \left[c_L\left(t\right) - c_I\left(t\right)\right] \\
&V_L \dot{c_L}\left(t\right)=-p_L \left[c_L\left(t\right) -c_I\left(t\right) \right] + p_G \left[c_G\left(t\right) - c_L\left(t\right) \right]
\label{}
\end{align}
where the dot signifies a time derivative.  Combining the two differential equations we obtain
\begin{equation}
\alpha \ddot{c_I}\left(t\right)+ \beta \dot{c_I}\left(t\right) + \gamma c_I\left(t\right) = \delta
\label{}
\end{equation}
where $\alpha, \beta, \gamma, \delta$ are constants obtained from parameters of the model.  The general solution of this equation is 
\begin{equation}
c_I(t) = C_1 e^{-\kappa_1 t}+C_2 e^{-\kappa_2 t} + C_3 \label{westerman}
\end{equation}
$c_I$ is then taken to be proportional to firing rate.  This equation allows adaptation to be modelled as a sum of two exponentials with different time constants.  The constants $C_1, C_2, C_3$, are fitted parametrically to adaptation data at different intensities \cite{westerman1984rapid,westerman1988diffusion}.  For the ZBC model, the parameterization is more complicated but the idea remains the same \cite{zhang2001phenomenological,heinz2001auditory}.

The second example is a ``universal'' phenomenological model based on the idea that mechanisms of adaptation can be reduced to a single current gated by another variable obeying a first order differential equation \cite{benda2003universal}.  The equations governing the firing frequency are given by
\begin{align}
& f=f_0\left(I-A \right) \\
& \tau_a  \dot{A} = -A+A_\infty\left(f\right)
\label{}
\end{align}
where $f$ is the firing rate, $A$ the gating variable and $f_0$ a function of $I-A$.  If $f_0$ is a linear function of its argument then $f$ can be solved for constant $I$ to be
\begin{equation}
f\left(t \right) = \left(f_0-f_\infty \right)e^{-t/\tau_\text{eff}} + f_\infty \label{benda1}
\end{equation}
Use of this equation requires empirical determination of both the peak growth curve $f_0$ and the steady-state growth curve $f_\infty$ as a function of stimulus intensity.  Both are fitted to the equation
\begin{equation}
\frac{f^\text{max}-f^\text{min}}{1+\exp\left[-k\left(I-I^0\right)\right]}+f^\text{min} \label{benda2}
\end{equation}
where $f^\text{min}, f^\text{max}, k, I^0$ are parameters to be determined from data fitting.
Equations (\ref{benda1}) and (\ref{benda2}) allow for the firing rate to be calculated as a function of intensity and duration.  Equation (\ref{benda2}) has been applied to the study of auditory interneurons in crickets but not to primary units \cite{hildebrandt2009origin}.  

The third example is a ``minimal'' phenomenological auditory firing rate model which can account for the transduction of intensity through a three stage process \cite{peterson2020phase,peterson2021simplified}.  The first stage connects the normalized current inside the mechanoelectrical transducer channel of an inner hair cell $M(t)$ to the instantaneous sound pressure level $P(t)$:
\begin{equation}
M(t)=\frac{1}{1+(1/M_0-1)e^{-b P(t)}} \label{heil1}
\end{equation}
The second stage involves the application of a low-pass filter to extract out only the DC component $M_{DC}$.  In the final stage, the firing rate is obtained by exponentiating $M_{DC}$ to obtain
\begin{equation}
R=R_{spont} \; e^{D M_{DC}} \label{heil2}
\end{equation}
The free parameters of the model include $M_0$, $b$ and $D$.  $R_{spont}$ is the level of spontaneous activity in the unit.  These equations are shown to give a good fit for both the peak firing rate (PR) and the steady-state firing rate (SS) as a function of intensity.  However, there is no explicit model to account for how the parameters vary as a function of adaptation.

The task of fitting (\ref{westerman}), (\ref{benda1}) or (\ref{heil2}) to adaptation data becomes an exercise in curve-fitting a surface over time and intensity.  Either the lower or the upper bound of (\ref{inequality}) can be used to reduce the total number of fitting parameters.  For the WS model, if we have determined empirically that the data of a particular fibre follows either the geometric or the arithmetic mean law we have an additional equation of either $\sqrt{\left(C_1+C_2+C_3\right) \text{SR}}=C_3$ or $\left(C_1+C_2+C_3+\text{SR}\right)/2=C_3$ to constrain the parameters in (\ref{westerman}).  A similar process can be adopted to constrain the values in either (\ref{benda1}) and (\ref{benda2}) or (\ref{heil1}) and (\ref{heil2}).  Alternatively, a second approach is discussed in the next section involving the slope of the firing rate function which can also be used to constrain the parameters.

\subsection{Implications for the Meddis model}
The three models considered thus far do not provide a mechanism for deriving the full dependency of firing rate on both intensity and duration.  In the final example, we consider the approach of Meddis  \cite{meddis1986simulation,meddis1988simulation,meddis1990implementation} which accounts for both changes in intensity and time by modelling the process of transduction through a set of non-linear compartmental equations governing the release of neurotransmitter in the auditory synapse \cite{meddis1986simulation}.  The Meddis model forms the heart of the approach in both the MAP and DSAM computational systems.  In this case, the dependency on the various concentrations is given by
\begin{align}
&\dot{q}(t)=y\left[M-q(t)\right]+x \; w(t)-k(t)q(t) \label{meddis1} \\ 
&\dot{c}(t)=k(t)q(t)-(l+r) c(t) \label{meddis2} \\
&\dot{w}(t)=r\; c(t)-x\; w(t) \label{meddis3} 
\end{align}
where $c(t)$ is the cleft concentration of transmitter, which is proportional to the firing rate, and the permeability $k(t)$ is a monotonic function of the sound pressure level $s(t)$.  The parameters of (\ref{meddis1}-\ref{meddis3}) and their interrelationships are defined clearly in \cite{meddis1990implementation} with proposed values of $A=5$, $B=300$, $g=2000$, $y=5.05$, $l=2500$, $x=66.3$, $r=6580$ and $m=1.0$.  In later versions of this model, the concentrations were quantized and the process made stochastic \cite{sumner2002revised,sumner2003adaptation}.  However, this does not change the overall dynamics of the model.

In its original formulation, the relationship between $k(t)$ and $s(t)$ was modelled via a simple mathematical rectification \cite{meddis1986simulation}.  In later models, it required the evaluation of a biophysical model of the inner hair cell and the signal filtering that precedes it  \cite{sumner2002revised}.  However, since the inequality (\ref{inequality}) is not dependent on the value of intensity itself, all that matters is that the average value of $k$ is a monotonic increasing function of stimulus intensity.

Under certain conditions, these equations can be solved analytically.  For choices of $l$ and $r$ that are large, (\ref{meddis2}) mimics the properties of a low-pass filter.  As such, at constant intensity, if the system is driven at high frequencies the equations can be approximated with $k(t)$ being constant.  The equations can now be solved in the $s$-domain using the Laplace transform \cite{zhang2005analysis}.  Zhang et al went further by approximating these equations under the condition of $l,r \rightarrow \infty$.  This means that the time-constant $1/\left(l+r \right)$ in (\ref{meddis2}) governing the transient behaviour between $c$ and $q$ is effectively zero.  From here, we can now decouple $c$ from (\ref{meddis3}) and replace it instead with the equivalent expression for $q$.  This then gives rise to the simplified model
\begin{align}
&\dot{q}(t)=y\left[M-q(t)\right]+x\; w(t)-\bar{k} \;q(t) \label{meddisp1} \\ 
&\dot{c}(t)=\bar{k} \; q(t)-(l+r) c(t) \label{meddisp2} \\
&\dot{w}(t)=\bar{k} \; u \; q(t)-x\; w(t) \label{meddisp3} 
\end{align}
where $u=r/(l+r)$ and $\bar{k}$ replaces $k(t)$ with the average permeability \cite{zhang2005analysis}.  In the $s$-domain, we have
\begin{align}
Q(s) &= \frac{\left[q(0) s+yM\right]\left(s+x \right)+s x w(0)}{s \left(s+x \right)\left(s+\bar{k} +y \right)-\bar{k} s u x} \label{meddisq} \\
C(s) &= \frac{\bar{k}Q(s)+c(0)}{s+l+r} \label{cs}
\end{align}
where $q(0)$, $c(0)$ and $w(0)$ are the initial values.  If the system is initially in quiet and steady-state prior to stimulus onset, we obtain the initial values by setting the derivatives in (\ref{meddis1}-\ref{meddis3}) equal to zero to obtain $c(0)=\bar{k}_\text{sp}My/\left[\bar{k}_\text{sp}l+\left(l+r\right)y\right]$, $q(0)=c(0)\left(l+r\right)/\bar{k}_\text{sp}$ and $w(0)=c(0)r/x$ where $\bar{k}_\text{sp}$ is the permeability/rate of transmitter leakage in silence.  Finally, an inverse Laplace transform of (\ref{cs}) gives $c(t)$ which is proportional to the spike activity of the adaptation response.  

The inequality derived earlier can provide insights into the Meddis model.  To see this, we solve the equations for the three relevant quantities.  The spontaneous rate SR can be calculated from the initial conditions, and the steady-state activity SS can be obtained by applying the final value theorem to (\ref{cs}):
\begin{align}
c_\text{SR} &= \frac{\bar{k}_\text{sp} M y}{\bar{k}_\text{sp} l+\left(l+r\right)y} \label{csr} \\
c_\text{SS} &= \frac{\bar{k} M y}{\bar{k} l+\left(l+r\right)y} \label{css}
\end{align}
The main difficulty, however, lies in the evaluation of the peak rate $c_\text{PR}$.  This involves inverting (\ref{cs}) into the time domain and then differentiating to find the location of the maximum activity -- a tedious and complex process.  However, under approximation, we can replace (\ref{cs}) by $C(s) \approx \bar{k}Q(s)/\left(l+r\right)$ for small values of $\bar{k}$ and obtain by the initial value theorem
\begin{equation}
c_\text{PR} =\frac{\bar{k} M y}{\bar{k}_\text{sp} l+\left(l+r\right)y} \label{cpr}
\end{equation}

\begin{figure}[!h]
\begin{center}
\includegraphics[width=.4\textwidth]{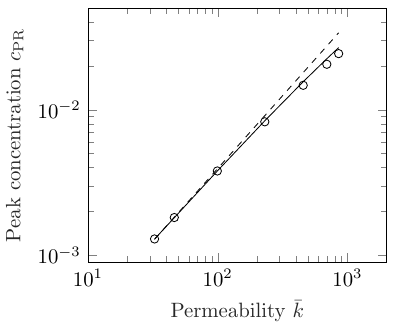}
\caption{Comparing the value of $c_{\text{PR}}$ evaluated by three methods.  Circles show the calculation by the original method in \cite{meddis1986simulation}; solid line was obtained by solving for the peak response from the inverse Laplace transform of (\ref{cs}); dashed line is from (\ref{cpr}).}\label{fig2}
\end{center}
\end{figure}

To test the validity of (\ref{cpr}) as an approximation to the true value, the peak activity $c_\text{PR}$ was evaluated by three methods.  The first method is to solve Meddis' original equations (\ref{meddis1}-\ref{meddis3}) using a sinusoidal input for $k(t)$ and following the technique outlined in \cite{meddis1986simulation}.  Since the location and value of peak activity depends on the exact phase of the signal, the solution of the differential equation was averaged across different values of the input phase, and the maximum value of $c(t)$ extracted. This constitutes a calculation of peak activity most consistent with Meddis' original model \cite{meddis1986simulation}.  The second method is to solve for $c(t)$ using the simplified Meddis model (\ref{meddisp1}-\ref{meddisp3}) with constant $\bar{k}$ and finding the maximum value.  Finally, $c_\text{PR}$ can be calculated directly from (\ref{cpr}).  In Fig. \ref{fig2}, equation (\ref{cpr}) can be seen to approximate the true value well with deviations observed only for larger $\bar{k}$.  As such, we are free to use (\ref{cpr}) as an approximation which becomes exact when $\bar{k}$ is small.

We are now in a position to see how the three derived quantities $c_\text{PR}$, $c_\text{SR}$ and $c_\text{SS}$ compare with the inequality.  Since the expression for the peak (\ref{cpr}) is only valid for small $\bar{k}$, we require an alternative form for (\ref{inequality}).  In Appendix B, the inequality is shown to be equivalent to the condition
\begin{equation}
\frac{dc_\text{SS}}{d \bar{k}}\bigg|_{\bar{k}=\bar{k}_\text{sp}}=\frac{1}{2}\frac{dc_\text{PR}}{d \bar{k}}\bigg|_{\bar{k}=\bar{k}_\text{sp}} \label{key}
\end{equation}
That is, the slope of $c_\text{SS}$ equals one half of the value of $c_\text{PR}$ for $\bar{k}$ near $\bar{k}_\text{sp}$.  Evaluating the derivatives from (\ref{css}) and (\ref{cpr}) and simplifying we obtain
\begin{equation}
\bar{k}_\text{sp}=\left(1+\frac{r}{l}\right)y \label{bridge}
\end{equation}
Equation (\ref{bridge}) is therefore the constraint imposed by the inequality on the model of Meddis.  Put in a different way: If the firing rate values obtained from the Meddis model are to match experimental data (which themselves obey the inequality), the model parameters must follow the constraint imposed by (\ref{bridge}).

An inspection of the parameters reported by Meddis appears to confirm this case.  In \cite{meddis1986simulation} and \cite{meddis1988simulation}, values of $r/l = 2.5$ and $y = 5$ were chosen as the parameters of best choice.  Independently, a value of $\bar{k}_\text{sp} = 33$ was used to give the correct level of spontaneous activity.  The two sides of (\ref{bridge}) are of the same order of magnitude.  These parameters were further refined in \cite{sumner2002revised,sumner2003adaptation} where $r/l$ remained 2.5 but $y$ was increased to 10.  

In its original conception, the search for a set of optimal parameter values for the Meddis model involved a laborious process of combing through a complex parameter landscape using the hill climbing method \cite{meddis1988simulation}.  The advantage conveyed by the inequality is that it allows for the derivation of (\ref{bridge}) which enables expedited exploration of the model by constraining the parameter space.  Equation (\ref{bridge}) would not be evident from an inspection of (\ref{meddis1}-\ref{meddis3}) alone.

\section{Methods}
All studies measuring adaptation at different intensities known to the author were included for analysis with the exception of \cite{rhode1985characteristics}.  Reasons for excluding this study included: (1) too few repetitions (5-10 trials per condition); (2) small bin width (100 $\mu$s); (3) inability to calculate firing rate consistent with reported data.  Additionally, for \cite{muller1991relationship}, figs. 3c and 3d were not included as the firing rate had yet to reach steady state.  

The data in Fig. \ref{fig1} were digitized and extracted from the original publications. The extracted data is available in the Supplementary Information.  For \cite{smith1975short}, peak PR and steady-state SS activities were obtained from fig. 4a, fig. 11a-d, fig. 12 and fig. 17a together with the spontaneous activity SR where available.  When SR was not reported, its value was set equal to the lowest measured value of SS (see panels e-g in Fig. \ref{fig1}).  For \cite{westerman1984rapid}, data was provided in terms of the coefficients $C_1,C_2,C_3$ of (\ref{westerman}).  See fig. 4 and 5 from their paper.  PR and SS were calculated from the equation and SR was provided for all units.  For \cite{yates1985very}, PR, SS and SR were extracted manually from the graphs.  For unit GP27/04, only two values are available as the peak rate for 15 and 20 dB exceeded the maximum value of the graph.  Similarly, PR, SS and SR were extracted from fig. 3 for \cite{fay1978coding}, fig. 3a and 3b for \cite{muller1991relationship}, and fig. 6 for \cite{sumner2012auditory}.  For \cite{peterson2021simplified}, values for PR and SS from figs. 12e-h were obtained using the fit to the data with their mathematical functions (\ref{heil1}) and (\ref{heil2}).  SR was reported for each unit.

\begin{acknowledgements}
This work was supported by a Discovery Grant from the Natural Sciences and Engineering Research Council of Canada (NSERC). 
\end{acknowledgements}

\section*{Conflict of interest}
The author declares no conflict of interest.

\appendix 
\section{Computer code to solve for adaptation response}
The following code was developed for the MATLAB programming environment (Mathworks 2023) but can be easily adapted to any other language to solve (\ref{gut1}-\ref{gut4}) numerically with only a few lines of code.  The parameters can be set to any positive value.    A forward Euler method is used to solve the differential equation in (\ref{gut4}).
\begin{verbatim}
% Numerical solution of the entropy equations (1-4)
% to a constant input beginning at t=2 (i.e. an 
% adaptation response to constant stimulus).
%
% Parameters can be set to any positive value.
% m is in steady-state and equals meq with i=0.
k=1; b=1; p=1; di=1; a=1; m=di^(p/2); i=0; 

% Time variable and increment
dt=0.1; t=0:dt:10; 

for j=1:length(t)

     % i=0 for t<2 and i=10 for t>=2
     if t(j)>2 i=10; end
     
     % Forward Euler method to solve ODE 
     m=m-dt*a*(m-(i+di)^(p/2));
     
     % Calculate response
     f(j)=.5*k*log(1+b*(i+di)^p/m);
     
end

% Plot calculations
plot(t,f); 
\end{verbatim}
While the above code was developed to solve for the adaptation response, a simple change to the line governing the stimulus allows for the code to solve the response to any time-varying input.  For the example illustrated in the code, note that $\text{SR}=0.35$, $\text{PR}=1.15$ and $\text{SS}=0.75$ satisfies the upper-bound of the inequality (\ref{inequality}).

\section{Slope of steady-state activity at low intensities}
There is an alternative way to express (\ref{inequality}): for low intensities the slope of steady-state activity SS with respect to $\bar{k}$ equals one half the slope of peak activity PR.  We will now prove this assertion.  Consider differentiable functions $f(x)$, $g(x)$ and $h(x)$ defined over the domain $x \ge 0$ satisfying the inequality $f(x) \le g(x) \le h(x)$ for all non-negative values of $x$.  Moreover, impose the conditions $f(0)=g(0)=h(0)=y_0$ and $f'(0)=h'(0)$.  $f'$ designates the first derivative of $f$ with respect to $x$, etc.  Clearly these statements are equivalent to $g'(0)=f'(0)=h'(0)$ by the definition of the derivative.

Next we recast the problem in terms of SS, PR and SR.  In this case, intensity replaces $x$ and $\text{GM}(0)=\text{SS}(0)=\text{AM}(0)=\text{SR}$ where AM and GM refer to the arithmetic and geometric means of SR and PR.  That is, at zero intensity both the GM and AM equals SR.  Moreover since SR is constant with respect to intensity, both $\text{AM}'(0)$ and $\text{GM}'(0)$ equals $\text{PR}'(0)/2$.  By the same steps as before, we conclude that $\text{SS}'(0)=\text{AM}'(0)=\text{GM}'(0)=\text{PR}'(0)/2$.  Therefore, the slope of SS equals one half the value of PR  at zero intensity, i.e. $\bar{k} =\bar{k}_\text{sp}$.

\bibliographystyle{spphys}
\bibliography{template1}   


\end{document}